\documentclass[reprint,amsmath,amssymb,aps,prd,]{revtex4-1}
\usepackage{graphicx}
\usepackage{epsfig,graphics,subfigure,psfrag,amsmath,amssymb}
\usepackage{tabularx}
\newcolumntype{C}{>{\centering\arraybackslash}X}
\usepackage{dcolumn}
\usepackage{bm}
\usepackage{overpic}
\usepackage{xspace}
\usepackage{rotating}
\usepackage{color}
\usepackage{multirow}
\usepackage{colortbl}
\usepackage{epstopdf}
\usepackage{float}
\usepackage{lineno}
\definecolor{aa}{RGB}{0,0,255}
\usepackage[bookmarksnumbered=true,colorlinks,urlcolor=aa,linkcolor=blue,anchorcolor=aa,citecolor=aa]{hyperref}

\newcommand{\PreserveBackslash}[1]{\let\temp=\\#1\let\\=\temp}
\newcolumntype{C}[1]{>{\PreserveBackslash\centering}p{#1}}
\newcolumntype{R}[1]{>{\PreserveBackslash\raggedleft}p{#1}}
\newcolumntype{L}[1]{>{\PreserveBackslash\raggedright}p{#1}}

\newcommand{\chicj}{\chi_{\textit{cJ}}}

\newcommand{\g}{\gamma}
\newcommand{\etap}{\eta^\prime}
\newcommand{\etcp}{\eta_{c}(2S)}

\newcommand{\pp}{\pi^+\pi^-}
\newcommand{\piz}{\pi^{0}}
\newcommand{\kk}{K^+K^-}
\newcommand{\kketap}{K^+K^-\eta^\prime}
\newcommand{\kkpipi}{K^+K^-\pi^+\pi^-}

\newcommand{\jpsi}{J/\psi}
\newcommand{\psip}{\psi(3686)}
\newcommand{\chicz}{\chi_{c0}}
\newcommand{\chico}{\chi_{c1}}
\newcommand{\chict}{\chi_{c2}}

\newcommand{\toetacp}{\psip \rightarrow \gamma\eta_{c}(2S) }

\newcommand{\detacp}{\eta_{c}(2S) \rightarrow K^{+}K^{-}\eta^\prime}
\newcommand{\dchicz}{\chi_{c0} \rightarrow K^{+}K^{-}\eta^\prime}
\newcommand{\dchico}{\chi_{c1} \rightarrow K^{+}K^{-}\eta^\prime}
\newcommand{\dchict}{\chi_{c2} \rightarrow K^{+}K^{-}\eta^\prime}
\newcommand{\dpipigam}{\eta^\prime \rightarrow \pi^{+}\pi^{-}\gamma}
\newcommand{\dpipieta}{\eta^\prime \rightarrow \pi^{+}\pi^{-}\eta}
\newcommand{\ra}{\rightarrow}
\newcommand{\gevcc}{{\rm GeV}/c^2}

\newcommand{\mevcc}{{\rm MeV}/c^2}
\newcommand{\mev}{\rm MeV}

\newcommand{\eff}{\epsilon}

\begin{document}
\graphicspath{{figure/}}
\DeclareGraphicsExtensions{.eps,.png,.ps}
\title{\boldmath Search for $\etcp$ $\ra$ $\kketap$ decay}
\author{
  \begin{small}
    \begin{center}
M.~Ablikim$^{1}$, M.~N.~Achasov$^{4,c}$, P.~Adlarson$^{75}$, O.~Afedulidis$^{3}$, X.~C.~Ai$^{80}$, R.~Aliberti$^{35}$, A.~Amoroso$^{74A,74C}$, Q.~An$^{71,58,a}$, Y.~Bai$^{57}$, O.~Bakina$^{36}$
% ...
\\
\vspace{0.2cm}
(BESIII Collaboration)\\
\vspace{0.2cm} {\it
$^{1}$ Institute of High Energy Physics, Beijing 100049, People's Republic of China\\
% ...
$^{3}$ Bochum  Ruhr-University, D-44780 Bochum, Germany\\
$^{4}$ Budker Institute of Nuclear Physics SB RAS (BINP), Novosibirsk 630090, Russia\\
% ...
$^{35}$ Johannes Gutenberg University of Mainz, Johann-Joachim-Becher-Weg 45, D-55099 Mainz, Germany\\
$^{36}$ Joint Institute for Nuclear Research, 141980 Dubna, Moscow region, Russia\\
% ...
$^{57}$ Southeast University, Nanjing 211100, People's Republic of China\\
$^{58}$ State Key Laboratory of Particle Detection and Electronics, Beijing 100049, Hefei 230026, People's Republic of China\\
% ...
$^{71}$ University of Science and Technology of China, Hefei 230026, People's Republic of China\\
% ...
$^{74}$ University of Turin and INFN, (A)University of Turin, I-10125, Turin, Italy; (B)University of Eastern Piedmont, I-15121, Alessandria, Italy; (C)INFN, I-10125, Turin, Italy\\
$^{75}$ Uppsala University, Box 516, SE-75120 Uppsala, Sweden\\
% ...
$^{80}$ Zhengzhou University, Zhengzhou 450001, People's Republic of China\\
\vspace{0.2cm}
$^{a}$ Deceased\\
% ...
$^{c}$ Also at the Novosibirsk State University, Novosibirsk, 630090, Russia\\
% ...
}\end{center}
    \vspace{0.4cm}
\end{small}
}
\affiliation{}


\begin{abstract}
Using $(2.712\pm0.014)\times10^{9}$ $\psi(3686)$ events collected with the BESIII detector operating at the BEPCII, we find an evidence of the 
$\eta_{c}(2S)\to K^+ K^- \eta^{\prime}$ decay with a statistical significance of 3.1$\sigma$. Its decay branching fraction is measured to be $(12.24\pm4.60(\mathrm{stat.})\pm2.37(\mathrm{syst.})\pm4.68(\mathrm{extr.}))\times 10^{-4}$, where the first uncertainty is statistical, the second is systematic, and the third uncertainty is from the branching 
fraction of the $\psi(3686)\to\gamma\eta_{c}(2S)$ decay. The upper limit on the product branching fraction 
$B[\psi(3686)\to\gamma\eta_{c}(2S)] \times$ $B[\eta_{c}(2S)\to K^+ K^- \eta^{\prime}]$ is set to be $1.14 \times 10^{-6}$ at 90\% confidence level. 
In addition, the branching fractions of $\chi_{c1}\to K^+ K^- \eta^{\prime}$ and $\chi_{c2}\to K^+ K^- \eta^{\prime}$ are updated to be 
$(8.47\pm0.09(\mathrm{stat.})\pm0.47(\mathrm{syst.}))\times 10^{-4}$ and $(1.53\pm0.04(\mathrm{stat.})\pm0.08(\mathrm{syst.}))\times 10^{-4}$, respectively. 
The precision is improved by twofold. 
\end{abstract}

\maketitle
\section{INTRODUCTION}
Charmonium states are composed of a pair of charm and anti-charm quarks. Their decay dynamics 
can be used to probe the strong interaction and to test models of Quantum Chromodynamics (QCD).
It is well predicted that the ratio of branching fractions of $\psip$ to $\jpsi$ decays into 
the same light hadron final state is around 13\%, the so-called ``12\% rule" which was first proposed 
by Appelquist and Politzer using perturbative QCD~\cite{intro1:qcd}. Although this rule has been verified 
for several hadronic channels~\cite{intro2}, there are exceptions for the $\rho\pi$ and 
$K^{*}\bar{K}$ final states, where the ratio is suppressed at least by an order of 
magnitude~\cite{intro3}, known as the ``$\rho\pi$ puzzle". 
The $\eta_{c}(2S)$ and $\eta_{c}(1S)$ are the spin-singlet partners of $\psip$ and $\jpsi$, 
respectively. The corresponding ratio for these two states was calculated in Refs.~\cite{intro4} and \cite{intro5}.
In Ref.~\cite{intro4}, the ratio is predicted to be 13\%, while the authors of Ref.~\cite{intro5} 
argue that the ratio is 1, considering that the spin-singlet states are dominantly decaying into light hadron 
final states via two-gluon intermediate states. 
Using the related experimental information from light hadron final states~\cite{pdg}, the authors 
of Ref.~\cite{in6:example} recently tested this branching fraction ratio in several decay 
modes~\cite{in6:example} and found that the experimental data significantly deviates from both 
theoretical predictions. 

The available data on the measurement of the branching fractions of the $\etcp$ decays is currently still limited. 
Although various experiments have searched for nineteen decay modes so far, only seven decay branching fractions 
have been measured~\cite{pdg}. Looking for new decay modes of $\etcp$ will provide a better understanding 
of its decay properties. Considering the radiative magnetic dipole transition,
$\psip\to\gamma\etcp$~\cite{in13:3c4c}, the $\etcp$ decays into $3(\pp)$~\cite{lsx}, 
$\pp\eta$~\cite{etacp-ppeta},  
and $K_{S}^{0} K\pi\pi\pi$~\cite{etacp-ppetac} 
were studied at BESIII. Obvious $\etcp$ signal events were observed in the $3(\pp)$~\cite{lsx} and $K_{S}^{0} K\pi\pi\pi$~\cite{etacp-ppetac} 
decay modes with statistical significances 
larger than $5\sigma$. 
The decay of $\etcp\to\kketap$ has not been measured yet, while the branching fraction of $\eta_c(1S)\to\kketap$ has been measured to be 
$(0.87\pm0.18)\%$~\cite{pdg}.

The $\chicj\ra\kketap$($J=1,2$) decays were first observed by BESIII in the $\psip\ra\gamma\chicj$ decay with 
$(106.4\pm0.9)\times10^{6}$ $\psip$ events. The branching fractions of $\chi_{c1}\ra\kketap$ and $\chi_{c2}\ra\kketap$ 
were measured to be $(8.75\pm0.87)\times10^{-4}$ and $(1.94\pm0.34)\times10^{-4}$, respectively~\cite{szt}. 
 
In this paper, we search for the $\detacp$ decay via the $\toetacp$ decay, 
and report the updated measurements of the branching fractions of $\chicj\ra\kketap$ ($J=1$ and $2$). 
The decay of $\dchicz$ is forbidden by spin-parity conservation.

\section{BESIII DETECTOR, DATA SAMPLE, AND MONTE CARLO SIMULATION}

The BESIII detector~\cite{bes3:detector} records symmetric $e^+e^-$ collisions provided by 
the BEPCII storage ring~\cite{bes3:detector2} in the center-of-mass (c.m.) energy ($\sqrt{s}$) 
range from 1.84 to 4.95~GeV, with a peak luminosity of $1 \times 10^{33}\;\text{cm}^{-2}\text{s}^{-1}$ 
achieved at $\sqrt{s} = 3.77\;\text{GeV}$. BESIII has collected large data samples in this 
energy region~\cite{bes3:detector3,EcmsMea,EventFilter}. 
The cylindrical core of the BESIII detector covers 93\% of the full solid angle and consists of 
a helium-based multilayer drift chamber~(MDC), a plastic scintillator time-of-flight system~(TOF), 
and a CsI(Tl) electromagnetic calorimeter~(EMC), which are all enclosed in a superconducting 
solenoidal magnet providing a 1.0~T magnetic field. The solenoid is supported by an octagonal 
flux-return yoke with resistive plate counter muon identification modules interleaved with steel. 
The charged-particle momentum resolution at $1~{\rm GeV}/c$ is $0.5\%$, and the ${\rm d}E/{\rm d}x$ 
resolution is $6\%$ for electrons from Bhabha scattering. The EMC measures photon energies with 
a resolution of $2.5\%$ ($5\%$) at $1$~GeV in the barrel (end cap) region. The time resolution 
in the TOF barrel region is 68~ps, while that in the end cap region was 110~ps. The end cap TOF 
system was upgraded in 2015 using multigap resistive plate chamber technology, providing a time resolution of 60~ps~\cite{bes3:detector4}, which benefits 83\% of the data used in this analysis.

$(2.712\pm0.014)\times10^{9}$ $\psip$~\cite{liucheng} events collected by the BESIII detector 
are used to search for $\etcp/\chicj\ra\kketap$ signal events, including $(107.7\pm0.6)\times10^{6}$ 
$\psip$ events collected in 2009, $(345.4\pm2.6)\times10^{6}$ $\psip$ events collected in 2012, 
and $(2259.3\pm11.1)\times10^{6}$ $\psip$ events collected in 2021. 
An additional continuum data set recorded at $\sqrt{s}=3.65$ GeV with an integrated 
luminosity of 401 $\rm{pb}^{-1}$~\cite{liucheng} is used to determine the nonresonant 
continuum background contributions.

Simulated samples produced with a {\sc geant4}-based~\cite{bes3:g4} Monte Carlo (MC) package 
which includes the geometric description of the BESIII detector and the detector response, 
are used to determine the detection efficiency and to estimate the backgrounds. The production 
of the $\psip$ resonance is simulated with the {\sc kkmc} generator~\cite{bes3:kkmc} taking 
into account the beam energy spread and initial state radiation in the $e^{+}e^{-}$ annihilations. 
Subsequent decays of the $\psip$ are modeled with {\sc evtgen}~\cite{bes3:evtgen} using 
branching fractions either taken from the Particle Data Group~\cite{pdg}, when available, 
or otherwise estimated with {\sc lundcharm}~\cite{bes3:lundcharm}. 
Final state radiation (FSR) from charged final state particles is incorporated using the 
{\sc photos} package~\cite{bes3:photon}.

In this analysis, exclusive MC simulations of the signal reactions $\toetacp$, $\gamma\chi_{c1,2}$ 
with $\etcp$, $\chi_{c1,2}\ra\kketap$ are generated, where the $\etap$ is reconstructed via 
its two main decay modes $\dpipigam$ and $\dpipieta$ with $\eta\ra\g\g$. The radiative 
transition $\toetacp$, $\gamma\chi_{c1,2}$ is modeled by following the angular distribution of 
$(1+\lambda \rm{cos}^{2}\theta)$, where $\theta$ is the polar angle of the radiative photon 
in the rest frame of the $\psip$ state with $\lambda$ set to be 1 for $\etcp$ and -1/3, 1/13 
for $\chico$ and $\chict$, respectively~\cite{dec-etacp-chicj}. The dynamics in the 
$\chicj\ra\kketap$ decay is considered by using Dalitz plots as input for the MC generator. 
For the $\dpipigam$ decay, a model that takes into account the $\rho-\omega$ interference and 
box anomaly is used~\cite{modelgpipi}. All other decays are modeled evenly distributed in 
phase space with {\sc evtgen}~\cite{bes3:evtgen}. An inclusive MC simulation is used to study
the background contributions, which includes the production of the $\psip$ resonance, the ISR 
production of the $\jpsi$, and the continuum processes incorporated in {\sc kkmc}~\cite{bes3:kkmc}.
An additional exclusive MC simulation of the reaction $\psip\ra\kketap$ is generated with 
{\sc evtgen}~\cite{bes3:evtgen} to better investigate background contribution of this process.

\section{EVENT SELECTION}
Charged tracks detected in the MDC are required to be within a polar angle ($\theta$)
range of $|\cos\theta|<0.93$, where $\theta$ is defined with respect to the $z$-axis,
which is the symmetry axis of the MDC. The distance of the closest approach to the 
interaction point must be less than 10~cm along the $z$-axis, $|V_{z}|$, and less than 
1~cm in the transverse plane, $|V_{xy}|$. The helix parameters of the charged tracks
in MC simulation are corrected to improve the consistency between data and MC 
simulation~\cite{helix}. 

Photon candidates are identified using showers in the EMC. The deposited energy of each 
shower must be more than $25~\mev$ in both the barrel ($|\cos\theta|<0.8$) and end cap
($0.86<|\cos\theta|<0.92$) regions. 
To suppress electronic noise and energy depositions unrelated to the event,  the difference 
between the EMC cluster time and the reconstructed event start time is required to be within 
$[0,~700]$~ns.

Charged particle identification (PID) is based on combining the d$E$/d$x$ and TOF information 
to construct a $\chi^{2}_{\rm PID}$. The values $\chi^{2}_{h-\mathrm{PID}}(i)$ are calculated 
for each charged track $i$ for each particle hypothesis $h$ ($h=\pi$ or $K$). If there is 
no valid PID information for the charged track, the $\chi^{2}_{h-\mathrm{PID}}(i)$ value is 
set to $0$. 

For the $\dpipigam$ mode, four charged tracks with net charge zero and at least two good 
photon candidates are required. A four-constraint (4C) kinematic fit is performed to constrain
the four-momenta of the final state particles to the initial one. The best photon candidates
and the species of the charged tracks in the final state are determined with the minimum 
$\chi^2_{\rm tot}=\chi^2_{\rm 4C}+\sum_{i=1}^{4}\chi^2_{h-\mathrm{PID}}(i)$, where $\chi^2_{\rm 4C}$
is given by the 4C kinematic fit and $\chi^2_{h-\mathrm{PID}}(i)$ is taken from PID. 
To suppress background events and to select $\dpipigam$ candidates, only the events satisfying 
$\chi^{2}_{\rm{4C}} < 25$ and $0.94~\gevcc < M_{\pi^{+}\pi^{-}\gamma} < 0.97~\gevcc$ are kept 
for the further analysis. Here, $M_{\pi^{+}\pi^{-}\gamma}$ denotes the invariant mass of $\pi^{+}\pi^{-}\gamma$. 
If both $\pi^+\pi^-\gamma$ combinations can meet the selection requirements, both are retained. 
The fraction of events with two entries is about 1.6\% in data. 

For the $\dpipieta$, $\eta\ra\gamma\gamma$ mode, four charged tracks with net charge zero and 
at least three good photon candidates are required. 
The photon pair passing a 1C kinematic fit is regarded as $\eta$ candidate, in which the 
$\gamma\gamma$ invariant mass, $M_{\gamma\gamma}$, is constrained to the nominal mass of 
$\eta$~\cite{pdg}. If there is more than one $\eta$ candidate in the event, all possible 
combinations are kept. Similarly, a 4C kinematic fit is performed, and the best photon 
candidate, $\eta$ candidate, and the species of the charged tracks in the final state 
are determined with the minimum $\chi^{2}_{\rm tot}$ = $\chi^{2}_{\rm 1C}$ + $\chi^{2}_{\rm 4C}$ + $\sum_{i=1}^{4}$ $\chi^{2}_{h-\mathrm{PID}}(i)$. 
To suppress background contributions from other decays and to select the $\eta\ra\gamma\gamma$ 
and $\dpipieta$ candidates, only the events satisfying $\chi^{2}_{\rm{4C}} < 20$, 
$0.52~\gevcc < M_{\gamma\gamma} < 0.57~\gevcc$, and 
$0.93~\gevcc < M_{\pi^{+}\pi^{-}\eta} < 0.98~\gevcc$ are kept for the further analysis, 
where $M_{\pi^{+}\pi^{-}\eta}$ is the invariant mass of the $\pi^{+}\pi^{-}\eta$ system.

The requirements of $\chi^{2}_{\rm 4C}$ in these two modes are optimized taking the maximum 
of the figure-of-merit (FOM) defined as $S/\sqrt{S+B}$, where $S$ and $B$ are the expected 
yields of the $\etcp$ signal and background events, respectively, in the $\etcp$ signal region defined 
as (3.60, 3.70) $\gevcc$. $S$ is calculated based on an assumed branching fraction of 
$B(\detacp)=B(\etcp\ra\kk\eta)=5\times 10^{-3}$~\cite{pdg}. 
$B$ is taken from the inclusive MC sample and scaled to data according to the total number of 
$\psip$ event. The mass window requirements for $\eta$ and $\etap$ candidates correspond to 
three times the mass resolution determined via fitting the corresponding invariant mass distributions 
$M_{\g\g}$ and $M_{\pi\pi\eta,~\g}$ from data with a sum of two Gaussian functions. 

Special requirements are placed to suppress various background contributions, the details are 
listed in Table~\ref{veto12}. 
For the $\dpipigam$ mode, the background events from $\psip\ra K^{+} K^{-}\pi^{+}\pi^{-}\piz/\eta$
are excluded by requiring $M_{\gamma\gamma}$ to be outside the $\piz$ and $\eta$ mass windows.
The background events from $\psip\ra\gamma\chicj$ 
($\chicj\ra\gamma\jpsi, \jpsi\ra K^{+} K^{-}\pi^{+}\pi^{-}$ or $\chicj\ra K^{+} K^{-}\pi^{+}\pi^{-}$) 
are excluded by requiring the invariant mass of the $\kkpipi$ system ($M_{\kkpipi}$) to be 
outside the $\jpsi$ and $\chicj$ ($J=0,1,2$) mass windows. The background events from 
$\psip\ra\pi^{+}\pi^{-}\jpsi$ are excluded by requiring the recoil mass of the $\pp$ pairs 
($M_{\pp}^{\rm recoil}$) to be outside the $\jpsi$ mass window. 
The background events from $\psip\ra\etap\phi,~\phi\ra\kk$ are excluded by requiring the invariant
mass of the $\kk$ system ($M_{\kk}$) to be outside the $\phi$ mass window. 
These requirements are named $\piz$ veto, $\eta$ veto, $\jpsi$ veto, $\chicj$ ($J=0,1,2$) veto, $\pp\jpsi$ 
veto, and $\phi$ veto, respectively. 

\begin{table}[htbp] 
	\caption{The mass windows used to veto background contributions.}
	\begin{center}    
		\begin{tabular}{ c  c  }  
			\hline\hline
\multicolumn{2}{c}{$\dpipigam$ mode}\\\hline
~Veto        ~&~ Mass window ($\gevcc$) \\\hline
$\piz$       ~&~ $0.122 < M_{\gamma\gamma} < 0.146$  \\
$\eta$       ~&~ $0.526 < M_{\gamma\gamma} < 0.566$  \\
$\jpsi$      ~&~ $3.08 < M_{\kkpipi} < 3.12$  \\
$\chicj$($J=0$)     ~&~ $3.37 < M_{\kkpipi} < 3.43$  \\
$\chicj$($J=1$)     ~&~ $3.48 < M_{\kkpipi} < 3.51$  \\
$\chicj$($J=2$)     ~&~ $3.53 < M_{\kkpipi} < 3.54$  \\
$\pp\jpsi$   ~&~ $3.091 < M_{\pp}^{\rm recoil} < 3.103$  \\
$\phi$       ~&~ $M_{K^{+}K^{-}} < 1.03 $  \\\hline\hline
\multicolumn{2}{c}{$\dpipieta$ mode}\\\hline
~Veto        ~&~ Mass window ($\gevcc$) \\\hline
$\piz(\gamma\gamma_{\eta})$ ~&~ $0.118 < M_{\gamma\gamma_{\eta}} < 0.150$  \\
$\eta\jpsi$   ~&~ $3.070 < M_{\eta}^{\rm recoil} < 3.133$  \\
$\phi$        ~&~ $M_{K^{+}K^{-}} < 1.03 $  \\\hline\hline

		\end{tabular}
	\end{center}
	\label{veto12}
\end{table}

For the $\dpipieta$ mode, the background events from $\psip\ra\kkpipi\piz$ are excluded by 
requiring $M_{\gamma\gamma_{\eta}}$ to be outside the $\piz$ mass window. Here $\gamma_{\eta}$ 
refers to the photon used to reconstruct $\eta$. 
The background events from $\psip\ra\eta\jpsi$ are excluded by requiring the recoil mass of 
$\eta$ ($M_{\eta}^{\rm recoil}$) to be outside the $\jpsi$ mass window. 
The background events from $\psip\ra\etap\phi,~\phi\ra\kk$ are excluded by requiring $M_{\kk}$ 
to be outside the $\phi$ mass window. These are named $\piz$ veto, $\eta\jpsi$ veto, and 
$\phi$ veto, respectively. 

After above selection and by using the signal MC simulations, the signal detection efficiencies 
for the $\dchico$, $\dchict$, and $\detacp$ signal channels are ($17.7\pm0.1$)\%, ($19.7\pm0.1$)\%, 
and ($14.3\pm0.1$)\% for the $\dpipigam$ mode, and ($9.9\pm0.1$)\%, ($10.9\pm0.1$)\%, and ($8.9\pm0.1$)\% for the $\dpipieta$ mode. 

\section{BACKGROUND ESTIMATION}
The background contribution is studied with the inclusive MC sample and the continuum data. 
In the $\dpipigam$ mode, about $99.7\%$ of the background events come from processes without 
$\etap$ as an intermediate state (non-$\etap$ processes). In the $\dpipieta$ mode, the dominant 
background contribution is also from the non-$\etap$ processes, which is about $90.1\%$ of the background 
events. Among the remaining background processes, $\psip\ra\kketap$ events are 
accumulated close to the $\etcp$ signal and are treated separately. 

\subsection{$\psip\ra\kketap$ background}
\begin{figure}[h] 
 \setlength{\abovecaptionskip}{0 cm}  
 \setlength{\belowcaptionskip}{0 cm}
 \centering
 \includegraphics[width=0.5\textwidth]{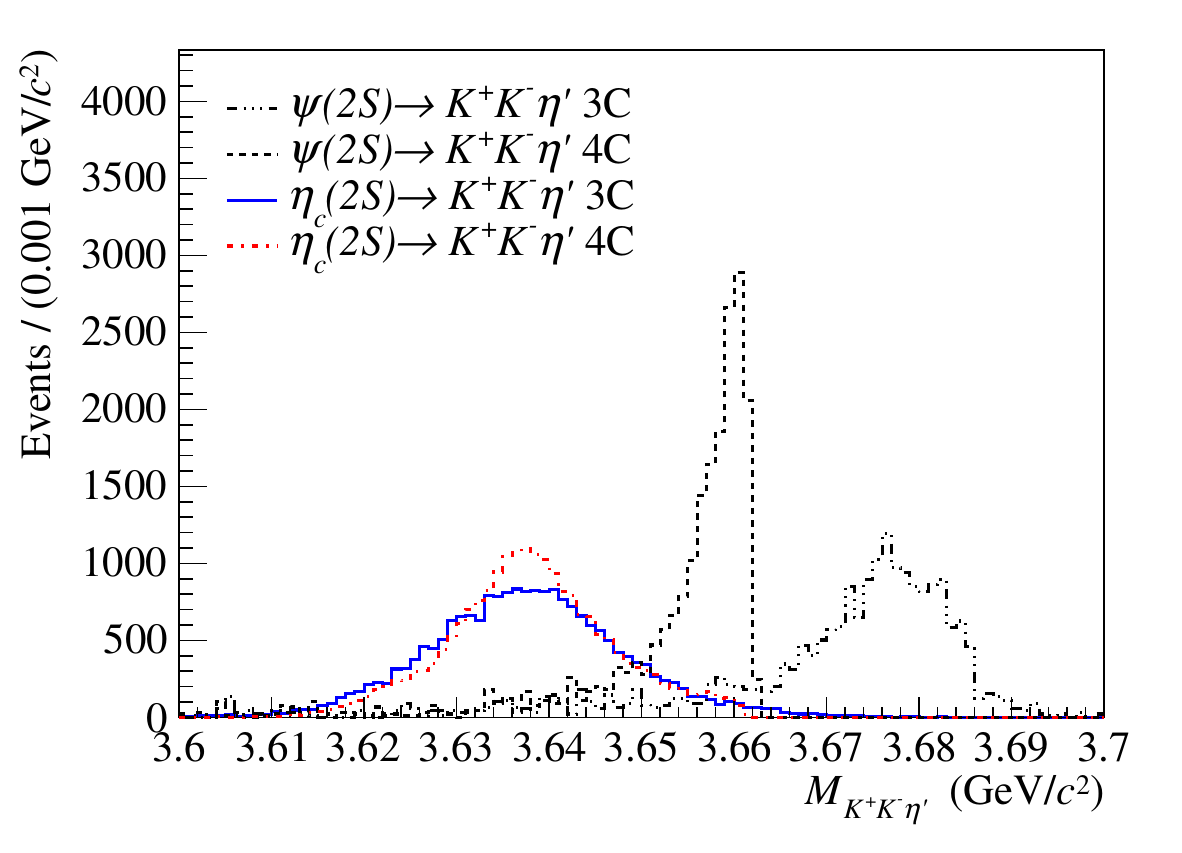}\\
 \caption{Comparison of the $M_{\kketap}$ distributions of the signal MC events from the 3C fit (blue solid line) and 4C fit (red dashed line), and those of the exclusive MC simulation of $\psip\ra\kketap$ from the 3C fit (black dashed-dotted line) and 4C fit (black dashed line). }
 \label{3c4c-kketap}
\end{figure}

Using the 4C kinematic fit information for the final state particles, background contributions 
from $\psip\ra\kketap(\gamma_{\rm{FSR}})$ processes create a peaking structure close to the 
$\etcp$ mass in the invariant mass distribution of the $\kketap$ system, that also contaminates 
the $\etcp$ signal. In order to better differentiate between the $\etcp$ signal and 
the $\psip\ra\kketap(\gamma_{\rm{FSR}})$ background contributions, a 3C kinematic fit is performed,
in which the energy of the radiative photon is not used as input~\cite{in13:3c4c}. In 
Fig.~\ref{3c4c-kketap}, a comparison between the results of the 3C and 4C fits are presented 
for $M_{\kketap}$ showing that the peak from the $\psip\ra\kketap(\gamma_{\rm{FSR}})$ events 
is pulled towards the $\psip$ peak in the 3C fit. 

\begin{figure*}[htbp] 
\setlength{\abovecaptionskip}{0 cm}   
\setlength{\belowcaptionskip}{0 cm}
\centering
  \begin{overpic}[width=0.48\textwidth]{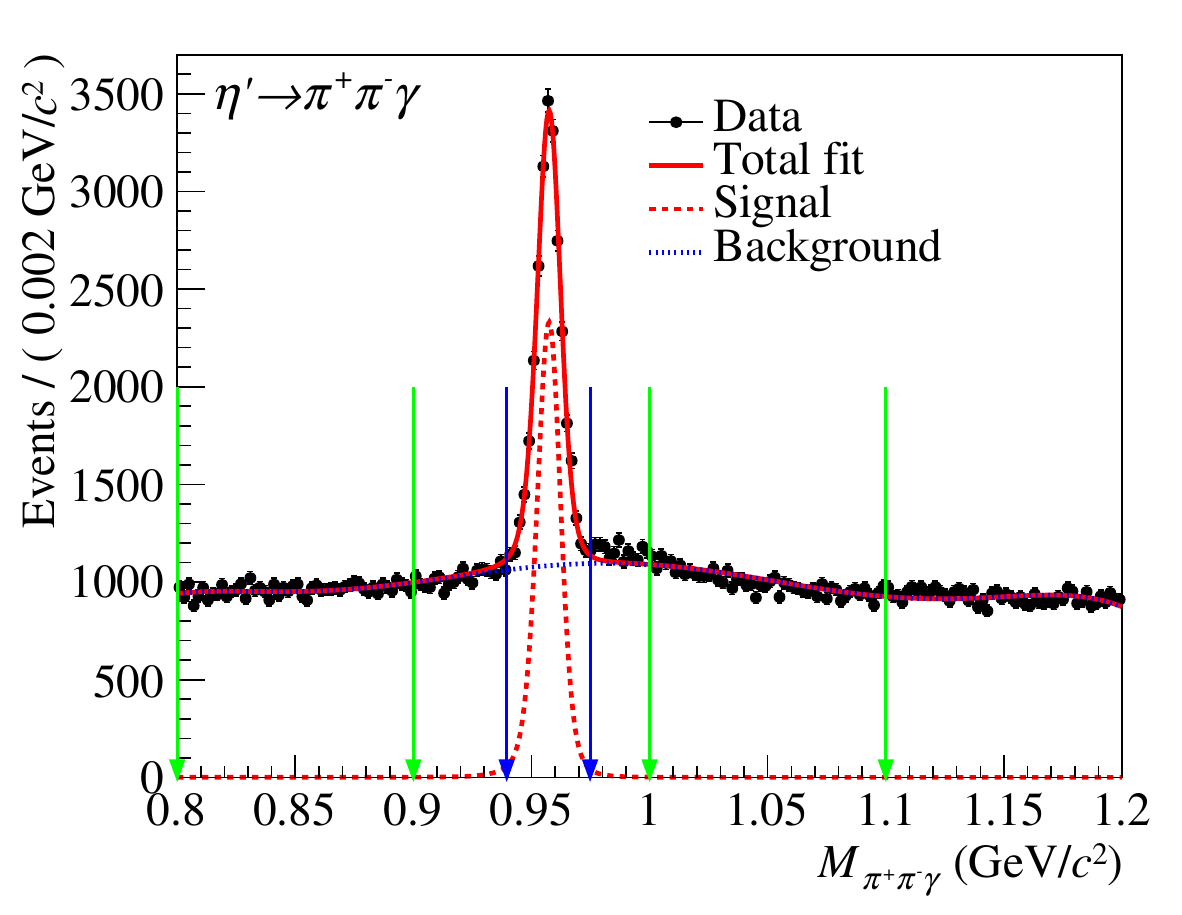}
  \put(50,1){(a)}
  \end{overpic}
  \begin{overpic}[width=0.51\textwidth]{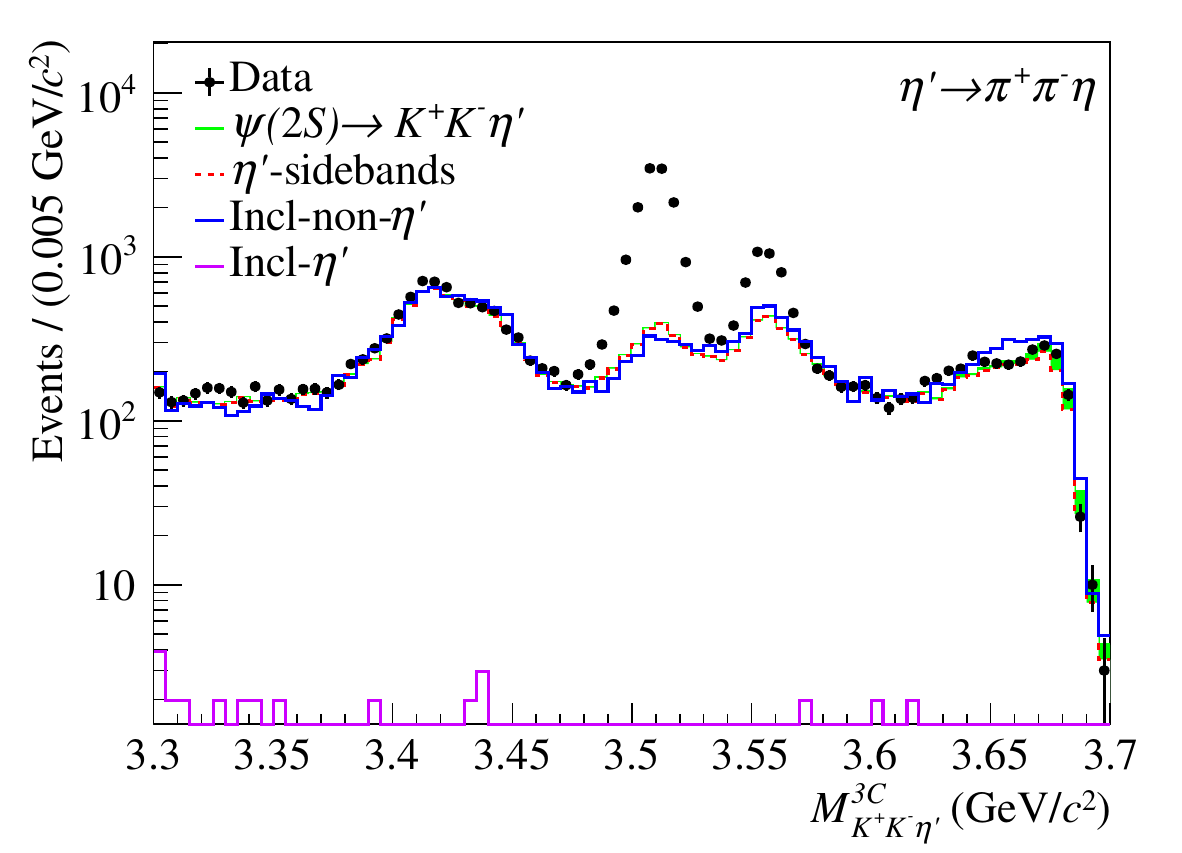}
  \put(50,1){(b)}
  \end{overpic}
  \centering
  \begin{overpic}[width=0.48\textwidth]{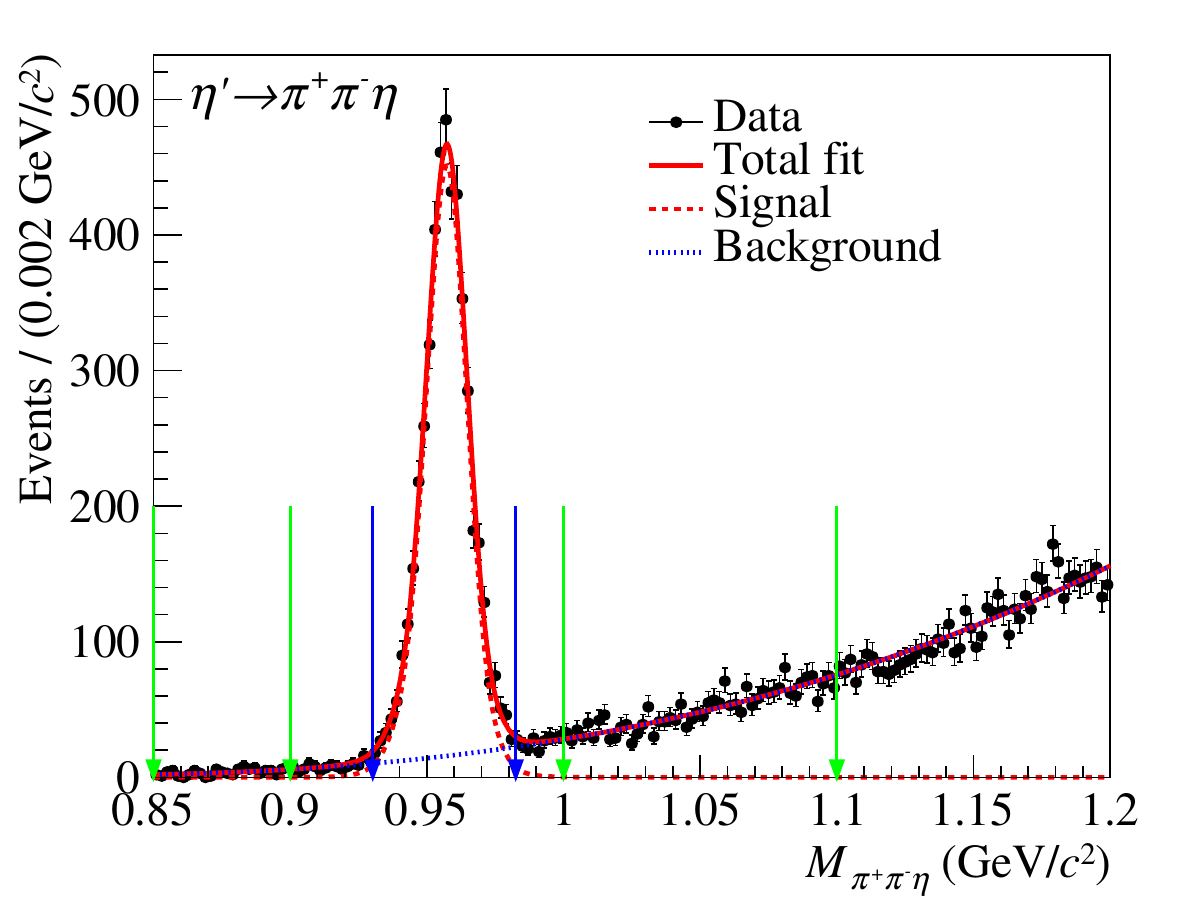}
  \put(50,0){(c)}
  \end{overpic}
  \begin{overpic}[width=0.51\textwidth]{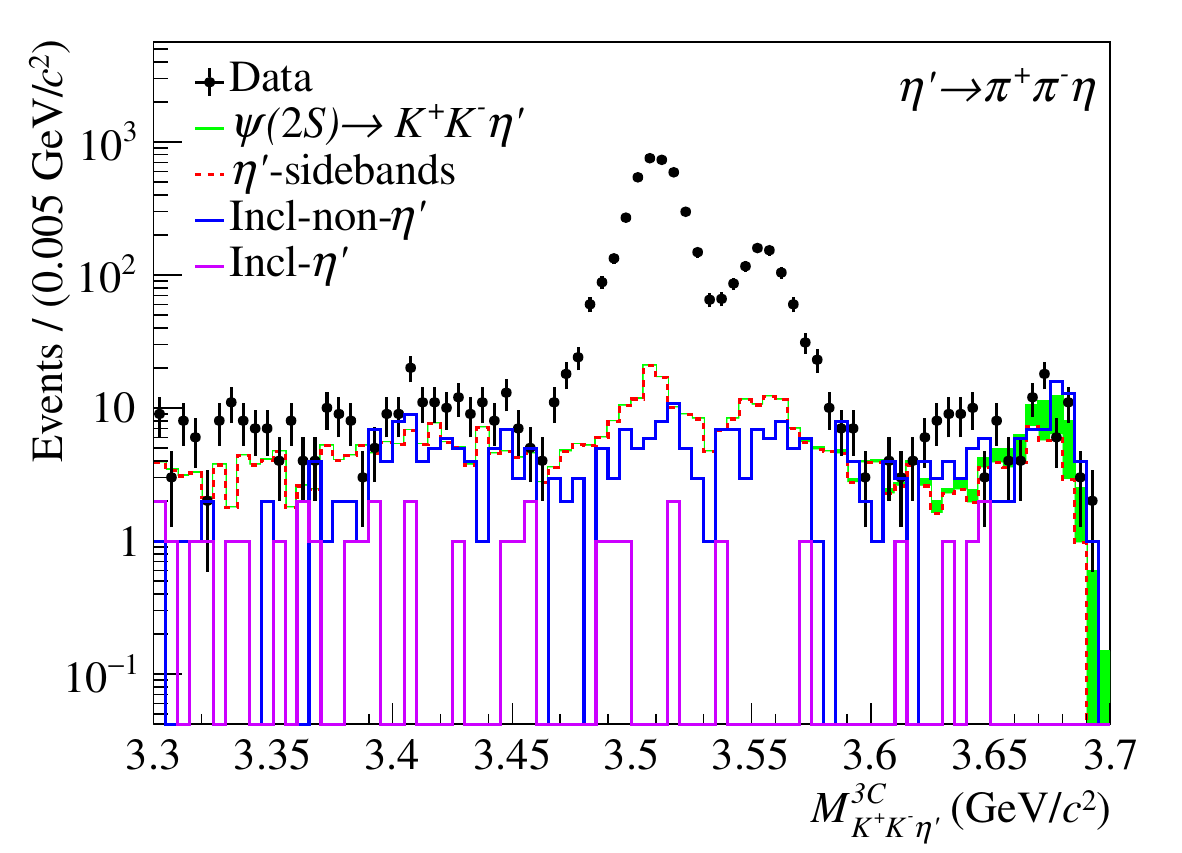}
  \put(50,0){(d)}
  \end{overpic}
  
 \caption{The left figures show fits to the (a) $M_{\pi^+\pi^-\gamma}$ and (c) $M_{\pi^+\pi^-\eta}$ distributions of the accepted candidate events in data (black dots with error bars). The red solid curves are the fit result, the dashed lines are the signal component, and the blue dashed lines represent the  background contribution. The pair of blue arrows indicate the $\etap$ signal region, and the two pairs of green arrows are the sideband regions. The right figures show comparisons of the $M_{K^+K^-\eta^\prime}^{\rm 3C}$ distributions between data and MC simulations for the $\dpipigam$ mode (b) and for the $\dpipieta$ mode (d). The red dashed histograms are the non-$\etap$ background events estimated with the $\etap$ sideband events in data, the blue solid histograms represent non-$\etap$ background events estimated with the inclusive MC sample, the purple solid histograms show the background contribution containing $\etap$ in the final state estimated using the inclusive MC sample, and the green filled histograms show the background contribution from $\psip\ra\kketap$ estimated using the MC simulation. The red dashed histograms and the green filled histograms are stacked.}
 \label{sideband}
\end{figure*}

The tail on the left side of the black dashed-dotted curve of the 3C case in Fig.~\ref{3c4c-kketap} for the background contribution 
is caused by FSR. The difference of the FSR ratio between data and MC simulation has been studied 
in a previous work using $\psip\ra\gamma\chicz,~\chicz\ra\pp\kk$~\cite{wyaqian}. There the FSR 
correction factor  
\begin{equation}\label{3}
\begin{aligned}  
 f_{\rm FSR} = R_{\rm FSR}^{\rm data} / R_{\rm FSR}^{\rm MC}
 \end{aligned}
\end{equation}
is determined to be $1.39\pm0.08\pm0.04$, where $R_{\rm{FSR}}^{\rm data}$ is the ratio of events 
with and without FSR in data, and $R_{\rm{FSR}}^{\rm{MC}}$ is the same ratio for MC simulations. 

The fraction of MC events with and without FSR is corrected with $f$, and the line shape of
$M_{\kketap}^{\rm{3C}}$ from the corrected MC sample, as shown in Fig.~\ref{3c4c-kketap}, is 
used in the fit to the $M_{\kketap}^{\rm 3C}$ distribution from data to extract the signal yield. 

\subsection{Non-$\etap$ background }
The non-$\etap$ background contribution is estimated using events in $\etap$ sideband regions. 
The $M_{\pp\gamma}$ distribution is fitted to determine the signal and sideband regions and 
the scale factor between the $\etap$ signal and sideband regions. 
The scale factor is calculated using the number of background events in the signal region 
and the sum of the number of background events in both sideband regions. These number of 
background events are determined using the background line shape with parameters fixed to 
the fitted values. 

For the $\etap\to\pp\gamma$ mode, the sideband regions are defined as $[0.8, 0.9]~\gevcc$ and 
$[1.0, 1.1]~\gevcc$. The scale factor is determined to be 0.194.
In the fit, the $\etap$ signal is described by an MC simulated shape convolved with a Gaussian 
function to account for differences in the detector resolution between data and MC, and the 
background is described by a sixth-order polynomial function. The fit result is shown in 
Fig.~\ref{sideband}(a). The fit quality, obtained by performing a $\chi^2$ test of the 
fitted curve and the binned mass spectrum, is $\chi^2/\mathrm{ndf}=247.2/190=1.3$. Here 
$\mathrm{ndf}$ is the number of degrees of freedom. 
There is a small and non-smooth background contribution on the right of the $\etap$ peak 
caused by the misselection of the photon. 
The effect on the sideband estimation is taken into account in the systematic uncertainty. 
For the $\dpipieta$ mode, the sideband regions are $[0.85, 0.90]~\gevcc$ and $[1.0, 1.1]~\gevcc$,
and the scale factor is 0.162. They are determined using the same method as for the $\dpipigam$ 
mode, except that the background is described by a third-order polynomial function. The fit 
result is shown in Fig.~\ref{sideband}(c) and the fit quality is $\chi^2/\mathrm{ndf}=137.7/168=0.8$. 

Figures~\ref{sideband}(b) and \ref{sideband}(d) show the comparison of the $M_{\kk\etap}^{\rm 3C}$ 
distributions from data together with selected background components estimated either with the $\etap$ sideband events from data, or with the inclusive and exclusive MC simulations. 
The comparison shows that the non-$\etap$ background events estimated from the $\etap$ sideband region (red dashed curve) can describe the non-$\etap$ background of inclusive MC (blue curve).

\subsection{Background including $\etap$}
The background contribution containing $\etap$ in the final state is about 0.3\% for the 
$\dpipigam$ mode and 9.9\% for the $\dpipieta$ mode, estimated using the inclusive MC sample. 
In Figs.~\ref{sideband}(b) and \ref{sideband}(d), the purple solid histograms show the 
$M_{\kk\etap}^{\rm 3C}$ distributions of this background contribution. They are smoothly 
distributed, thus could be fitted to a polynomial function.

\subsection{Continuum contribution}
\begin{figure}[htbp]  
\setlength{\abovecaptionskip}{0 cm}   
\setlength{\belowcaptionskip}{0 cm}
\centering
  \hbox{
  \begin{overpic}[width=0.50\textwidth]{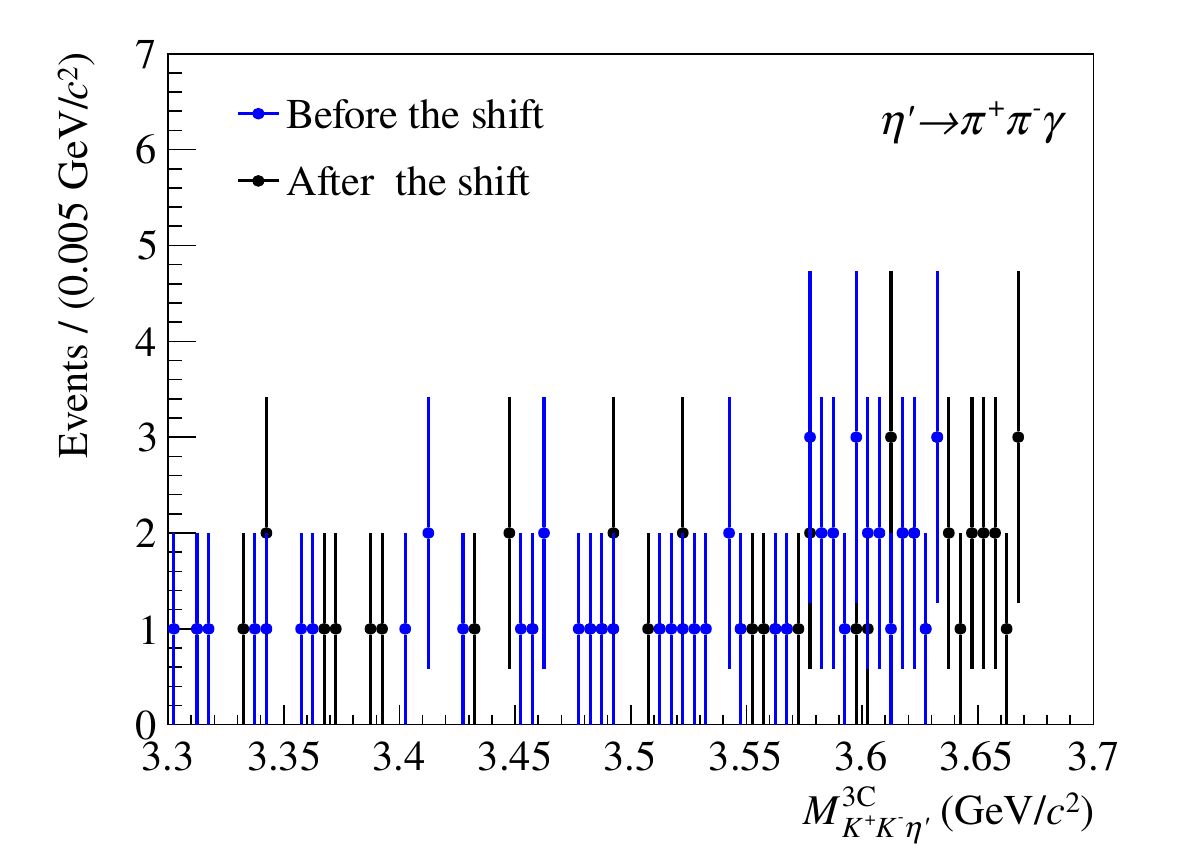}
  \put(50,1){(a)}
  \end{overpic}
  }
  \centering
  \hbox{
  \begin{overpic}[width=0.50\textwidth]{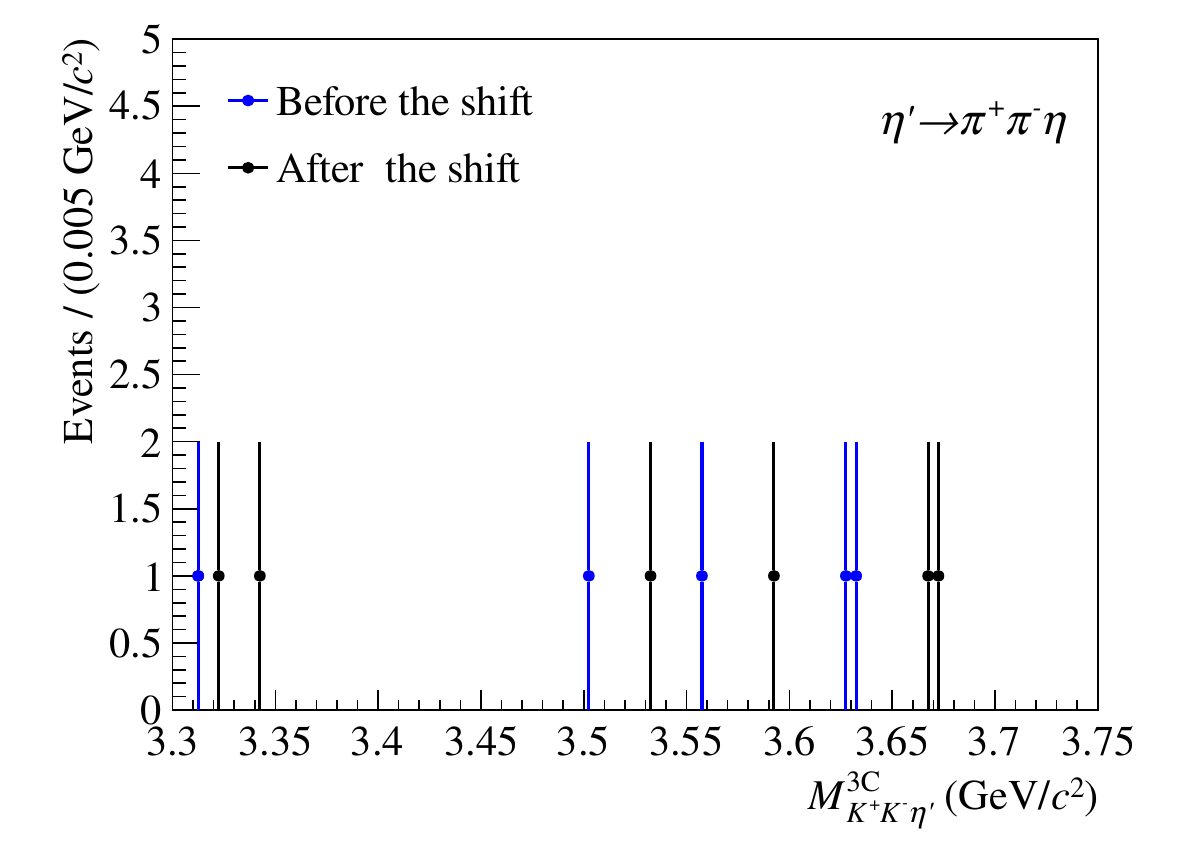}
  \put(50,0){(b)}
  \end{overpic}
  }
	\caption{The $M_{\kketap}^{\rm 3C}$ distributions for the (a) $\dpipigam$ and (b) $\dpipieta$ modes from the continuum data at 3.650 GeV. The blue dots with error bars are before the mass shift, the black dots with error bars are after the mass shift.  } 
	\label{continuum-shift}
\end{figure}

The background contribution from the continuum process is estimated using the continuum data 
set at $\sqrt{s}=$3.650~GeV. 
The event selections applied to this data sample are the same as for the $\psip$ data sample, 
except that the c.m. energy is changed from 3.686 GeV to 3.650 GeV in the kinematic fit. 
Considering the difference of the c.m. energies, the $M_{\kketap}^{\rm 3C}$ ($m$ in Eq.(2)) 
from the continuum
data is shifted and corrected by 
\begin{equation}\label{fitpdf-signal}
m_{\rm shifted} \ra a(m-m_{0}) + m_{0},
\end{equation}
where $m_{0}$ = 1.945 $\gevcc$ is the mass threshold for $\kketap$ which should not be shifted, 
$a=(3.686~\gevcc-m_{0})/(3.650~\gevcc-m_{0})=1.021$ is the shift coefficient for $\sqrt s = 3.650$ GeV to ensure that the events 
at 3.650 GeV are shifted to 3.686 GeV. The $M_{\kketap}^{\rm 3C}$
distributions before and after the shift are shown in Fig.~\ref{continuum-shift}. 

The mass spectrum of $M_{\kketap}^{\rm 3C}$ is normalized based on the differences in the 
integrated luminosity and cross section. For the continuum data set at 3.650 GeV, the scale 
factor is calculated via  
\begin{equation}
	f_{\rm continuum} = \frac{\mathcal{L}_{3.686}}{\mathcal{L}_{3.650}} \times \frac{\sigma_{3.686}}{\sigma_{3.650}}=8.31,
	\label{eq-con-scale}
\end{equation}
where $\mathcal{L}_{3.686} = 3.4~\mathrm{fb}^{-1}$ and $\mathcal{L}_{3.650} = 401~\mathrm{pb}^{-1}$~\cite{liucheng} 
are the integrated luminosities at 3.686 GeV and 3.650 GeV, respectively, and $\sigma$ is the 
production cross section, which is assumed to be proportional to $1/s$. 
The shape and the estimated continuum background yield of the continuum data sample with applied
scaling factor are used to fit the data. 

\begin{figure*}[htbp] 
\setlength{\abovecaptionskip}{0 cm}   
\setlength{\belowcaptionskip}{0 cm}
\centering
  \begin{overpic}[width=0.495\textwidth]{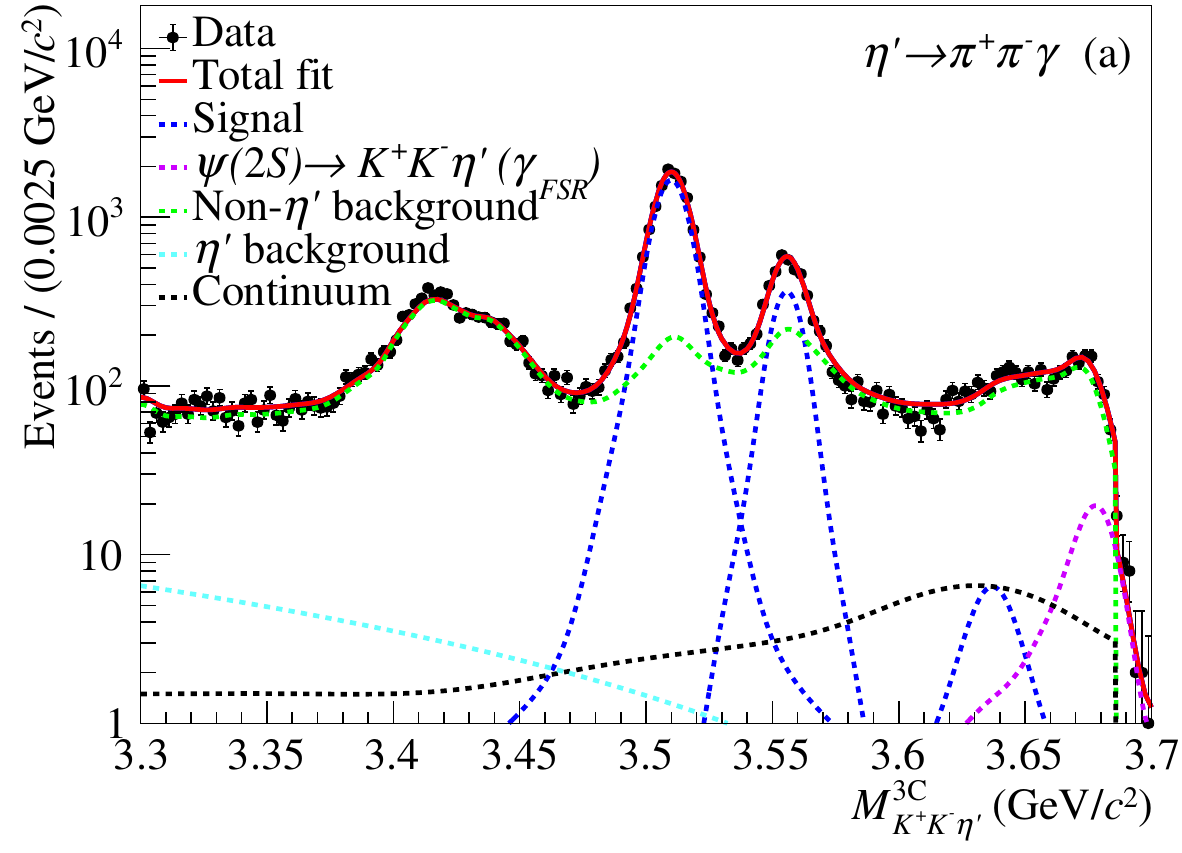}
  \put(50,1){}
  \end{overpic}
  \begin{overpic}[width=0.495\textwidth]{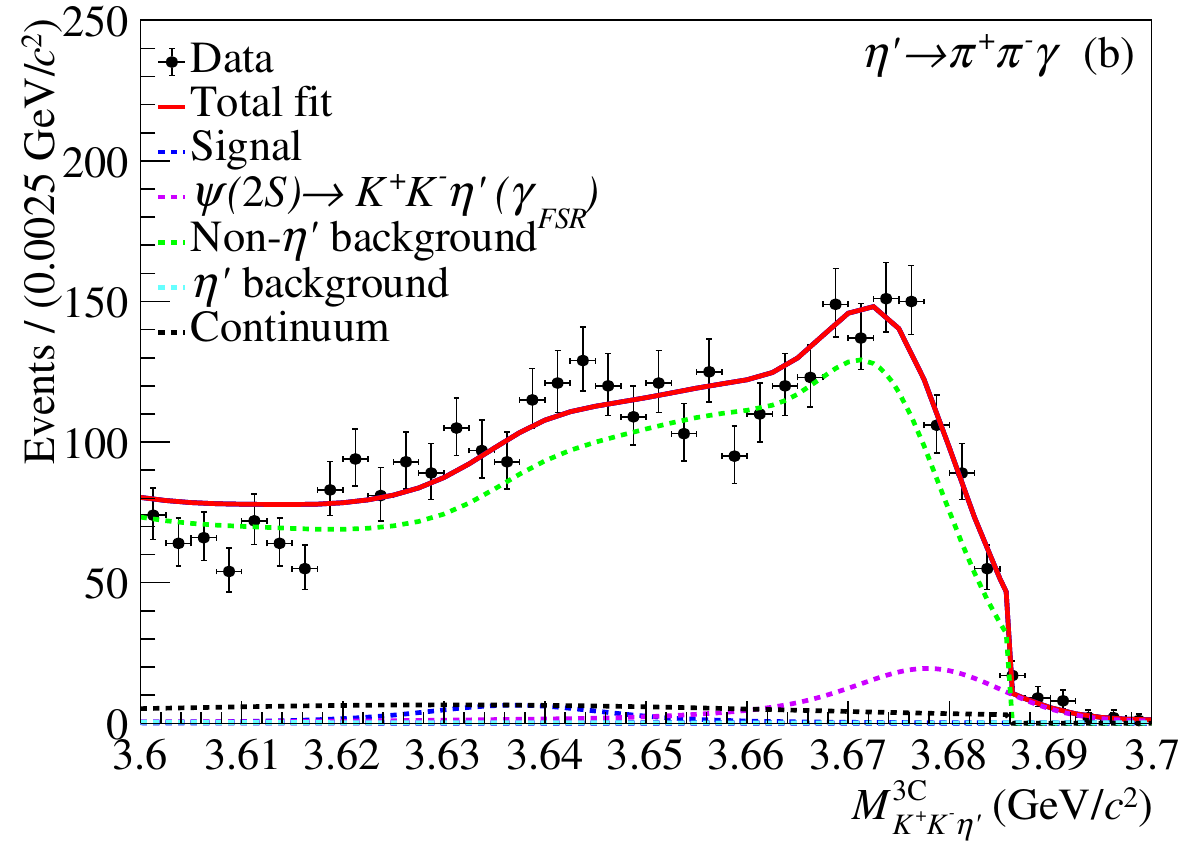}
  \put(50,1){}
  \end{overpic}
\centering
  \begin{overpic}[width=0.495\textwidth]{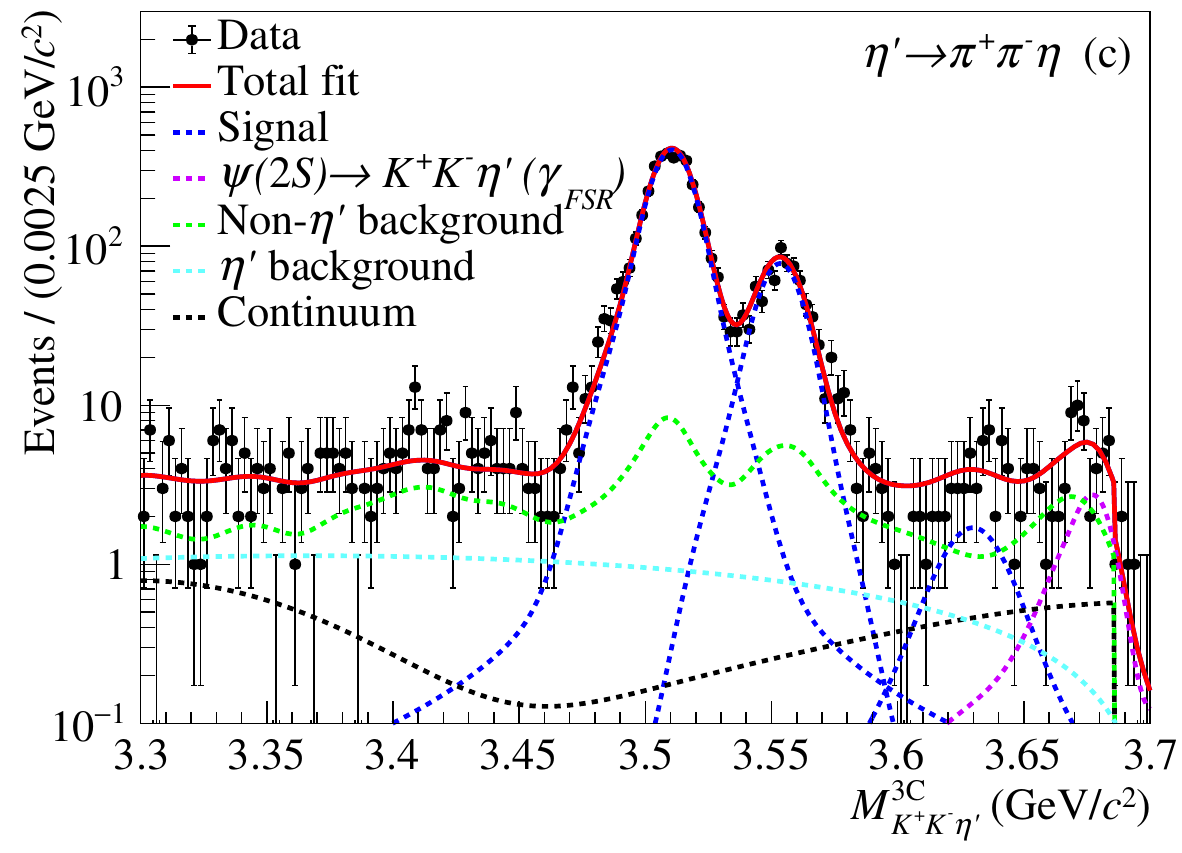}
  \put(50,0){}
  \end{overpic}
  \begin{overpic}[width=0.495\textwidth]{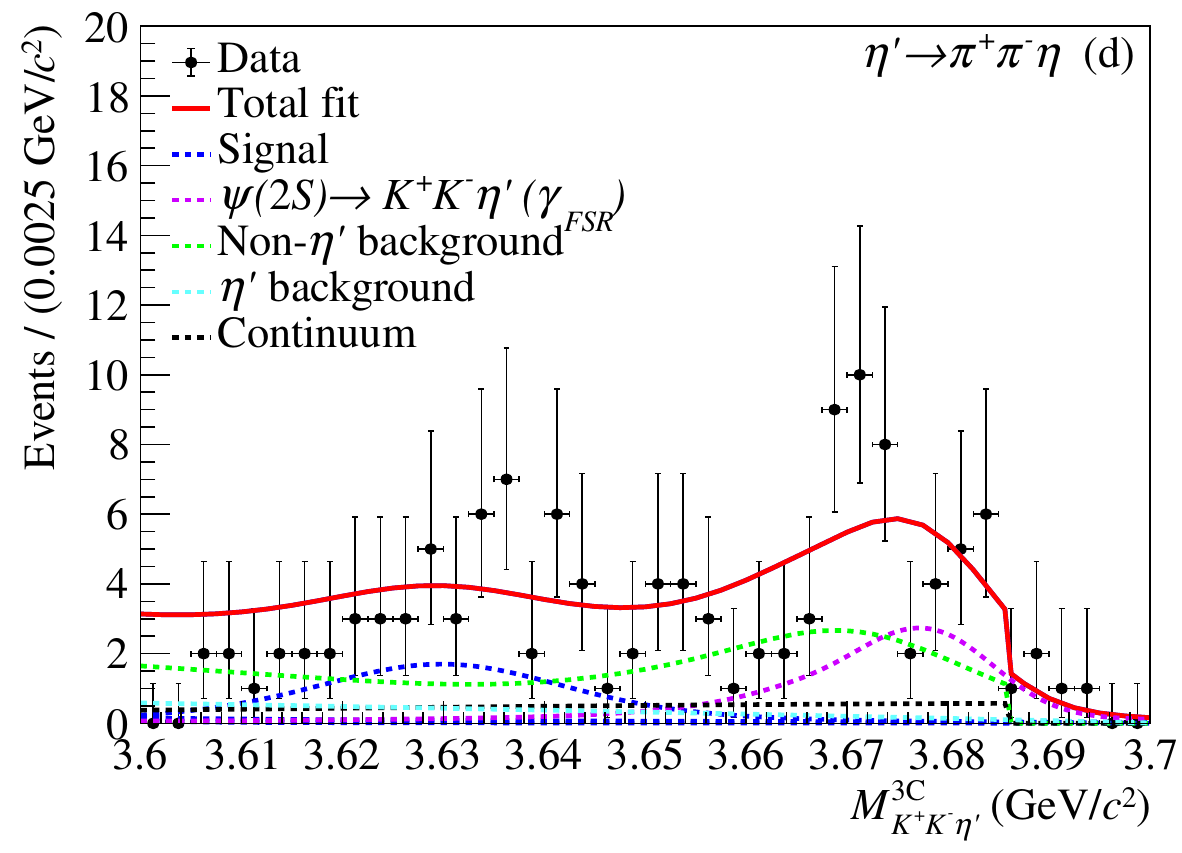}
  \put(50,0){}
  \end{overpic}

 \caption{The simultaneous fit to the $M_{\kketap}^{\rm 3C}$ distributions in the whole fit range in (a) and (c), and the range only containing $\etcp$ signal in (b) and (d). The dots with error bars are data, the red solid curves correspond to the best fit result, the blue dashed lines show the $\etcp$ and $\chicj~$($J=1,2$) signal shapes, the purple dashed lines represent the contribution from $\psip\ra\kketap$, the green dashed lines show the non-$\etap$ contribution from the $\etap$-sideand events from data, the black dashed lines are the continuum contribution, and the cyan dashed lines represent the remaining background.   }
 \label{simfit-to-data-12}
\end{figure*}

\section{SIGNAL EXTRACTION}
The signal yields are determined from an unbinned maximum likelihood fit to the 
$M_{\kketap}^{\rm 3C}$ spectra of the $\dpipigam$ and $\dpipieta$ mode simultaneously. The 
function used in the fit is defined as
\begin{equation}
\mathrm{PDF}_{\rm total} = N^{\rm sig} \cdot S + \sum_{i=1}^{4} N^{\rm bkg}_{i}\cdot B_{i}. 
\end{equation}
Here, $N^{\rm sig}$ and $N^{\rm bkg}_{i}$ represent the numbers of signal events and background 
events, and $S$ and $B_{i}$ are the signal and background PDFs. 

The signal PDF is given by 
\begin{equation}\label{fitpdf-signal}
    S=[(E_{\gamma}^{3}  \times f_{d}(E_{\gamma}) \times BW(m) )\otimes DG(m) \times \eff(m)] \otimes G(\delta m, \delta\sigma),
\end{equation}
where $m$ corresponds to the $M_{\kketap}^{3C}$ distributions after the 3C kinematic fit,   
$E_{\gamma}$ is the energy of the radiative photon in the rest frame of the $\psip$, 
and $BW(m)$ is the relativistic Breit-Wigner function used to describe the resonance. The mass 
and total width are fixed to the PDG values for $\etcp$~\cite{pdg} and are free parameters 
for $\chi_{c1,2}$. 
The factor $E_{\gamma}^{3}$  is introduced due to the energy dependence of the transition 
matrix element from the radiative photon energy in the radiative magnetic diplole transition 
process. 
The function $f_{d}(E_{\gamma})$ damps the diverging tail raised by $E_{\gamma}^{3}$. 
It is adapted from previous work by the KEDR Collaboration~\cite{kedr} with 
$f_{d}(E_{\g})=\dfrac{ E_{0}^{2} }{ E_{\gamma}E_{0} + (E_{\gamma} - E_{0})^{2} }$, 
where $E_{0} = \dfrac{M_{\psip}^{2} - M_{\etcp/\chicj}^{2}}{2M_{\psip}}$ is the most probable 
energy of the radiative photon. 
$DG(m)$ represents the mass resolution, determined by fitting 
$\Delta M = M_{\kketap}^{\rm 3C} - M_{\kketap}^{\rm true}$ with a sum of two Guassian functions. 
Here, $M_{\kketap}^{\rm 3C}$ is the invariant mass of $\kketap$ from the 3C kinematic fit and 
$M_{\kketap}^{\rm true}$ is the mass at generator level. 
$\eff(m)$ denotes the efficiency curve, which is obtained from the signal MC simulations and 
parameterized by a generalized ARGUS function~\cite{ARGUS}, defined as 
\begin{equation}\label{fitpdf-eff}
    \eff(m)=m(1-(\frac{m}{m_{0}})^{2})^{p}\times e^{c(1-(\frac{m}{m_{0}})^{2})},
\end{equation}
where $m_{0}=3.67~\gevcc$ is the kinematic limit (cut-off) corresponding to the $\etcp$ mass including 
detector resolution. The parameters $p$
and $c$ are determined to be $p=0.311, ~c=-2.74$ for $\dpipigam$ and $p=0.238, ~c=-2.04$ 
for $\dpipieta$.   
To account for the mass resolution difference between data and MC simulations, an additional Gaussian 
function, $G(\delta m, \delta\sigma)$ is included. For the $\chicj$ signals, the parameters of this 
Gaussian function are floated, while for the $\etcp$ signal they are fixed to the values extrapolated
from $\chicj$ with a linear assumption, which are 
$\delta m = (0.12\pm0.18)~\mevcc$ and $\delta \sigma = (-0.87\pm1.87)~\mevcc$ for the $\dpipigam$ mode,
$\delta m = (-6.52\pm2.73)~\mevcc$ and $\delta \sigma = (5.39\pm5.06)~\mevcc$ for the $\dpipieta$ mode. The mass and width of $\chi_{c1,2}$ determined from the fit are consistent with the results from PDG. 

In the simultaneous fit, the branching fraction of $\etcp/\chicj\ra\kketap$, $\mathcal{B}^{\rm sig}$ is 
the same for the two $\eta^\prime$ modes, it is connected to the number of signal events via 
\begin{equation}
 \mathcal{B}^{\rm sig}=\frac{N^{\rm sig}}{N_{\psip}\cdot\mathcal{B}(\psip\to\gamma\etcp/\chicj)\cdot\bar{\epsilon}},  
\end{equation}
where $\bar{\epsilon}=\mathcal{B}(\dpipigam)\cdot\epsilon_{1}+\mathcal{B}(\dpipieta)\cdot\mathcal{B}(\eta\to\gamma\gamma)\cdot\epsilon_{2}$ is the average detection efficiency, $\mathcal{B}$ is the 
branching fraction of the corresponding process, $\eff_{1}$ and $\eff_{2}$ are the signal detection 
efficiencies for the $\dpipigam$ and $\dpipieta$ modes, respectively. 

The background contributions have been well estimated in section IV. They are decomposed into 
four components: the $\psip\ra\kketap$ background ($B_{1}$), the non-$\eta'$ background ($B_{2}$), 
the continuum contribution ($B_{3}$), and the remaining backgrounds ($B_{4}$). The $B_{1}$ shape 
is taken from the $\psip\ra\kketap$ MC simulation, corrected by the FSR ratio $f$,  
the $B_{2(3)}$ line shape is directly determined from data, and $B_{4}$ is a second-order 
polynomial function. $N_{1}^{\rm bkg}$ and $N_{4}^{\rm bkg}$ are free parameters in the fit, while 
$N_{2}^{\rm bkg}$ and $N_{3}^{\rm bkg}$ are fixed.   
We added a truncation at $3.686~\gevcc$ to the line shape of the continuum sample and $\etap$-sideband to account for the kinematic boundary. 

\begin{table*}[htbp] 
	 \caption{Branching fractions for the three different signal reactions. The first uncertainty is statistical and the second is systematic. For $\detacp$, the third uncertainty is from $\mathcal{B}[\toetacp]$. }  
	\begin{center}    
		\begin{tabular}{ c  c c }  
			\hline\hline
			~Channel  ~&~ $\mathcal{B}~(\times 10^{-4})$~&~$\rm{PDG~}(\times 10^{-4})$\\\hline
$\detacp$ ~&~ $\rm 12.24\pm4.60(stat.)\pm2.37(syst.)\pm4.68(extr.)$     ~&~  -                   \\      
$\dchico$ ~&~ $\rm 8.47\pm0.09(stat.)\pm0.47(syst.)$                    ~&~ $8.8\pm0.9$         \\        
$\dchict$ ~&~ $\rm 1.53\pm0.04(stat.)\pm0.08(syst.)$                    ~&~ $1.94\pm0.34$
\\\hline\hline      
			
		\end{tabular}
	\end{center}
	\label{sim}
\end{table*}

\begin{table*}[htbp] 
	\caption{ Systematic uncertainties for the branching fraction measurements [$\%$]. For $\detacp$, the value in parentheses is the total systematic uncertainty without the uncertainty from $\mathcal{B}(\toetacp)$. }
	\begin{center}    
		\begin{tabular}{ c  c  c  c }  
			\hline\hline
			Source                                     ~&~ $\detacp$   ~&~ $\dchico$ ~&~ $\dchict$ ~~\\\hline
            \multicolumn{4}{c}{Multiplicative term}\\\hline
   		Total number of $\psip$ events             ~&~ 0.5        ~&~ 0.5      ~&~ 0.5  \\
			Tracking                                   ~&~ 4.0        ~&~ 4.0      ~&~ 4.0    \\
			Photon reconstruction                      ~&~ 2.3        ~&~ 2.3      ~&~ 2.3  \\
			$\mathcal{B}(\etap\ra\pp\gamma/\eta)$      ~&~ 1.5        ~&~ 1.5      ~&~ 1.5  \\
			$\mathcal{B}(\psip\ra\gamma\etcp/\chicj)$  ~&~38.2        ~&~ 2.5      ~&~ 2.1  \\
			MC statistics                              ~&~ 0.1        ~&~ 0.1      ~&~ 0.1  \\
			MC generator                               ~&~ -          ~&~ 0.6      ~&~ 0.0  \\ 
            Kinematic fit                              ~&~ 1.1        ~&~ 0.8      ~&~ 1.1  \\
			$\eta$ mass window                         ~&~ 0.0        ~&~ 0.0      ~&~ 0.0  \\
			$\etap$ mass window                        ~&~ 0.2        ~&~ 0.2      ~&~ 0.2  \\
			All background veto                        ~&~ 3.7       ~&~ 0.0      ~&~ 0.3  \\\hline
            \multicolumn{4}{c}{Additive term}\\\hline
                FSR factor                                 ~&~ 1.3        ~&~ 0.1      ~&~ 0.2  \\
			Number of continuum events                 ~&~ 3.8        ~&~ 0.0      ~&~ 0.0  \\
      Shape of continuum                         ~&~  13.4   ~&~    0.2   ~&~ 0.3   \\
      Shape of non-$\etap$ events                ~&~   4.3      ~&~  0.1  ~&~  0.2  \\
			Shape of $\psip\ra\kketap$                 ~&~ 5.2        ~&~ 0.0      ~&~ 0.1  \\
			Shape of other background                  ~&~ 0.2        ~&~ 0.2      ~&~ 0.1  \\
			Region of $\etap$ sideband                 ~&~ 4.4        ~&~ 0.1      ~&~ 0.2  \\
                Number of non-$\etap$ events               ~&~ 1.8        ~&~ 0.0      ~&~ 0.2  \\
			Damping function form                      ~&~ 2.9       ~&~  0.0      ~&~  0.2 \\
			First kind of resolution                   ~&~ 3.2        ~&~ 0.5      ~&~ 0.1  \\
			Second kind of resolution($\delta m$)      ~&~ 5.5        ~&~ -         ~&~ -     \\
			Second kind of resolution($\delta \sigma$) ~&~ 3.7        ~&~ -         ~&~ -     \\
			Efficiency curve                           ~&~ 1.3        ~&~ 0.2      ~&~ 0.3  \\
			Mass of $\etcp$                            ~&~ 3.1        ~&~ 0.1      ~&~ 0.1  \\
			Width of $\etcp$                           ~&~ 1.1        ~&~ 0.0      ~&~ 0.1  \\\hline
            Total        ~&~ 42.9 (19.4)~&~ 5.6      ~&~ 5.5  \\\hline\hline		
   
		\end{tabular}

	\end{center}
	\label{sys}
\end{table*}

The fit results are shown in Fig.~\ref{simfit-to-data-12}, and the fit quality is 
$\chi^{2}/\rm{ndf} = 228.2/145 = 1.6$. The results for the branching ratios are summarized in Table~\ref{sim}. 
The spectra from the two decay modes are also fitted separately, the results are consistent with each 
other. An input and output check is performed to check the bias of the fit procedure, the output values 
are consistent with the input ones. 
The statistical significance of the $\etcp$ signal is estimated to be 3.1$\sigma$, calculated using the 
difference of the logarithmic likelihoods~\cite{sigma}, -2ln$(\mathcal{L}_{0}/\mathcal{L}_{\rm max})$. 
Here $\mathcal{L}_{\rm max}$ and $\mathcal{L}_{0}$ are the maximized likelihoods with and without the 
$\etcp$ signal component, respectively. 
The difference in the number of degrees of freedom ($\Delta$ndf=1) has been taken into account. 

\section{SYSTEMATIC UNCERTAINTY ESTIMATION}
The systematic uncertainties in the branching fraction measurements are summarized in Table~\ref{sys}. 
They are classified in two categories: the multiplicative terms and the additive terms. 

The multiplicative terms refer to the uncertainties due to the total number of $\psip$ events, the detector efficiency, and the branching fractions. They are estimated separately for the two $\etap$ decay modes. Varying the corresponding values by $\pm 1\sigma$, the changes on the average detection efficiency is assigned as the systematic uncertainty. They are described in detail in the following. 
 
The total number of $\psip$ event is determined to be $(27.12\pm0.14)\times10^{8}$~\cite{liucheng}. Its uncertainty is 0.5\%. 

The uncertainty from the tracking is assigned to be 1\% per track using the control samples of $\jpsi\ra\piz\pp$ or $\jpsi\ra K_{S}^{0}K\pi+c.c$~\cite{sys-track}. 
The uncertainty from the photon reconstruction is studied using the control samples $\jpsi\ra\rho^{0}\pi^{0}$ and $e^{+}e^{-}\ra\gamma\gamma$~\cite{sys-photon}, and is assigned to be 1.0\% per photon. 
 
The uncertainties from the branching fractions of $\etap\ra\pp\gamma$, $\etap\ra\pp\eta$, and 
$\eta\ra\gamma\gamma$ are 1.4\%, 1.2\%, and 0.5\%, respectively. The uncertainties from the branching fractions of $\psip\ra\gamma\etcp$, $\chico$, and $\chict$ are 38.2\%~\cite{psiptogametacp}, 2.5\%, and 2.1\%~\cite{pdg}, respectively. 

The Dalitz plot, momentum, and cos$\theta$ distributions of $K^{+}$, $K^{-}$, and $\etap$ in the signal MC samples produced with the {\sc body3} model are basically consistent with the distributions in data. Therefore, the uncertainty from the MC generator is estimated by changing the number of bins of the input Dalitz plot. Due to limited statistics of the MC samples used, a 0.1\% uncertainty is taken for each decay. 
 
In the kinematic fit, the helix parameters of charged tracks in MC samples have been corrected to improve the consistency between data and MC simulations~\cite{helix}. Half of the differences of efficiencies with and without the helix parameter correction are taken as the systematic uncertainties. This is a conservative estimation, as the uncertainties of the correction factors are at 10\% level, and the difference between data and MC simulation
in the $\chi^2_{\rm 4C}$ distributions after the correction is much smaller than that before. 

For the uncertainty arising from the mass window requirements of $\eta$ and $\etap$, the $M_{\eta}$ and $M_{\etap}$ distributions from data are fitted with the corresponding simulated shapes convolved with a Gaussian resolution function. 
The $\eta$ and $\etap$ distributions from the signal MC samples are smeared with the resultant Gaussian function. The change of the signal detection efficiency is taken as the uncertainty. 

The uncertainties due to different mass vetoes are examined via a Barlow test~\cite{barlowtest}. A significant 
deviation ($\zeta$) between the nominal result and the systematic test is defined as  
\begin{equation}
   \label{zeta}
   \begin{aligned}
 &\zeta = \dfrac{|\mathcal{B}_{\rm nominal}^{\rm sig} - \mathcal{B}_{\rm test}^{\rm sig}|}{\sqrt{|\sigma_{\mathcal{B}_{\rm nominal}^{\rm sig}}^{2} - \sigma_{\mathcal{B}_{\rm test}^{\rm sig}}^{2}|}}, 
 \end{aligned}
\end{equation}
where $\sigma_{\mathcal{B}}$ is the statistical error of the branching fraction. We vary the veto mass windows 
used and repeat the simultaneous fit. The obtained $\zeta$ distribution shows no significant deviation. As a 
conservative estimation, the maximum difference in the branching fraction is taken as the systematic uncertainty.  
The systematic uncertainty caused by different vetoes estimated by the Barlow tests is mainly caused by statistical fluctuations. 
Therefore, for vetoes where there is no obvious trend and $\zeta$ 
is less than 2, the systematic uncertainty of $\etcp$ and $\chi_{c1,2}$ is set to be 0. 
Taking the $\piz$ veto in the $\dpipigam$ mode for example, we vary the requirement on $M_{\gamma\gamma}$ to be 
(0.110, 0.158), (0.112, 0.156), (0.114, 0.154), (0.116, 0.152), (0.118, 0.150), (0.120, 0.148), (0.122, 0.146), 
(0.124, 0.144) $\gevcc$. For other mass vetoes, the mass windows are changed using the similar method, as detailed 
in Table~\ref{veto12-sys}. 

\begin{table}[htbp] 
	\caption{The mass window variations (in unit of $\gevcc$) used to estimate the uncertainties for all mass vetoes.  }
	\begin{center}    
		\begin{tabular}{ c  c  c  c }  
			\hline\hline
\multicolumn{4}{c}{$\dpipigam$ mode}\\\hline
~Veto        ~&~ Lower limit ~&~ Upper limit ~&~ Step \\\hline
$\piz$       ~&~ (0.110, 0.124) ~&~ (0.144, 0.158) ~&~ 0.002 \\
$\eta$       ~&~ (0.518, 0.532) ~&~ (0.560, 0.574) ~&~ 0.002  \\
$\jpsi$      ~&~ (3.072, 3.086) ~&~ (3.114, 3.128) ~&~ 0.002 \\
$\chicj$($J=0$)     ~&~ (3.365, 3.372) ~&~ (3.428, 3.435) ~&~ 0.001  \\
$\chicj$($J=1$)     ~&~ (3.475, 3.482) ~&~ (3.508, 3.515) ~&~ 0.001  \\
$\chicj$($J=2$)     ~&~ (3.525, 3.532) ~&~ (3.538, 3.545) ~&~ 0.001  \\
$\pp\jpsi$   ~&~ (3.086, 3.093) ~&~ (3.101, 3.108) ~&~ 0.001  \\
$\phi$       ~&~        -       ~&~ (1.028, 1.035) ~&~ 0.001  \\\hline\hline
\multicolumn{4}{c}{$\dpipieta$ mode}\\\hline
~Veto        ~&~ Lower limit ~&~ Upper limit ~&~ Step \\\hline
$\piz(\gamma\gamma_{\eta})$ ~&~ (0.114, 0.121) ~&~ (0.147, 0.154) ~&~ 0.001 \\  
$\eta\jpsi$   ~&~ (3.060, 3.074) ~&~ (3.129, 3.143) ~&~ 0.002  \\
$\phi$        ~&~        -       ~&~ (1.028, 1.035) ~&~ 0.001  \\\hline\hline

		\end{tabular}
	\end{center}
	\label{veto12-sys}
\end{table}

The additive terms are related to the determination of $N_{\rm sig}$ from the fit. Different fit conditions are tested in the simultaneous fit, and the difference on the branching fraction is taken as systematic uncertainty. These are introduced in the following. 

The uncertainty due to the FSR is estimated by varying the factor of $f_{\rm{FSR}} = 1.39\pm 0.08 \pm 0.04$~\cite{wyaqian} by $\pm 1\sigma$. 

The uncertainty from the number of continuum events introduced by the $\sigma$ in Eq.~\eqref{eq-con-scale} is estimated by changing $1/s$ to $1/s^{n}$, where $n$=1.5 or 3. Here the value of $n$ is taken from Ref.~\cite{zhaojingyi}, where several $e^+e^-\to light~hadron$ processes are measured, and the dependency of the cross section on $s$ varies from 1.5 to 3. The number of continuum events is fixed to the values estimated using the new scale factor, the changes on the branching fraction is taken as uncertainty. 
 
In the nominal fit, the line shapes of $B_{1,2,3}$ are extracted from MC simulated sample or data samples using RooKeysPdf~\cite{keyspdf}. The uncertainty from the corresponding line shape is estimated by replacing it with RooHistPdf~\cite{histpdf}. As for $B_{4}$, it is changed from a second-order polynomial function to the line shape extracted from the background events including $\etap$ from the inclusive MC sample using RooKeysPdf. 

The uncertainty due to the scale factor of the non-$\etap$ background events is estimated by replacing the parameters of the background line shape with alternative values, which are obtained from a multi-dimensional Gaussian sampling. In the sampling, the covariance matrix from the fit to $M_{\pi^+\pi^-\gamma}$ ($M_{\pi^+\pi^-\eta}$) is used as input.
A total of 10000 samplings are performed, thereby giving 10000 different scale factors and numbers of non-$\etap$ background events. A Gaussian fit is performed on this distribution, the obtained standard deviation of the Gaussian distribution is regarded as the uncertainty of the number of non-$\etap$ background events. The number of non-$\etap$ background events is varied by $\pm1\sigma$ in the simultaneous fit to estimate the systematic uncertainty. Additional systematic uncertainty from non-$\etap$ background events is considered in the $\dpipigam$ mode by changing the right sideband region from [1.0, 1.1]$~\gevcc$ to [1.002, 1.098]$~\gevcc$ and [0.998, 1.102]$~\gevcc$, given the fit shown in Fig.~\ref{sideband}(a) is not perfect on the right side of the $\etap$ peak. 

The uncertainty from the damping function is evaluated with an alternative damping function used by the CLEO~\cite{cleo} Collaboration, $f_{d}(E_{\gamma}) = \exp(-E_{\gamma}^{2} / 8\beta^{2})$. 
The uncertainty from parametrizing the efficiency curve is estimated by changing the ARGUS function into RooHistPDF. The uncertainty from the fixed mass and width of $\etcp$ is estimated by varying them by $\pm 1 \sigma$. 
To estimate the uncertainty from the parameterization of the mass resolution of signal, we use RooKeysPdf to replace the double Gaussian function. The systematic uncertainty from the mass resolution difference between data and MC simulations is considered by varying $\delta m$ and $\delta\sigma$ by $\pm 1\sigma$ for $\etcp$. There is no such term for $\chico$ and $\chict$, since the parameters of the Gaussian function are free parameters in the fit. 

\section{Upper limit of the branching fraction}
Since the significance of $\etcp\to\kk\etap$ is just above $3\sigma$, we also determine the upper limit on 
the product branching fraction of $B[\psip\ra\gamma\etcp] \times B[\etcp\ra\kketap]$ at 90\% confidence level 
(C.L.) using a Bayesian method~\cite{bayesian}.

\begin{figure}[h]  
 \setlength{\abovecaptionskip}{0 cm}  
 \setlength{\belowcaptionskip}{0 cm}
 \centering
 \includegraphics[width=0.50\textwidth]{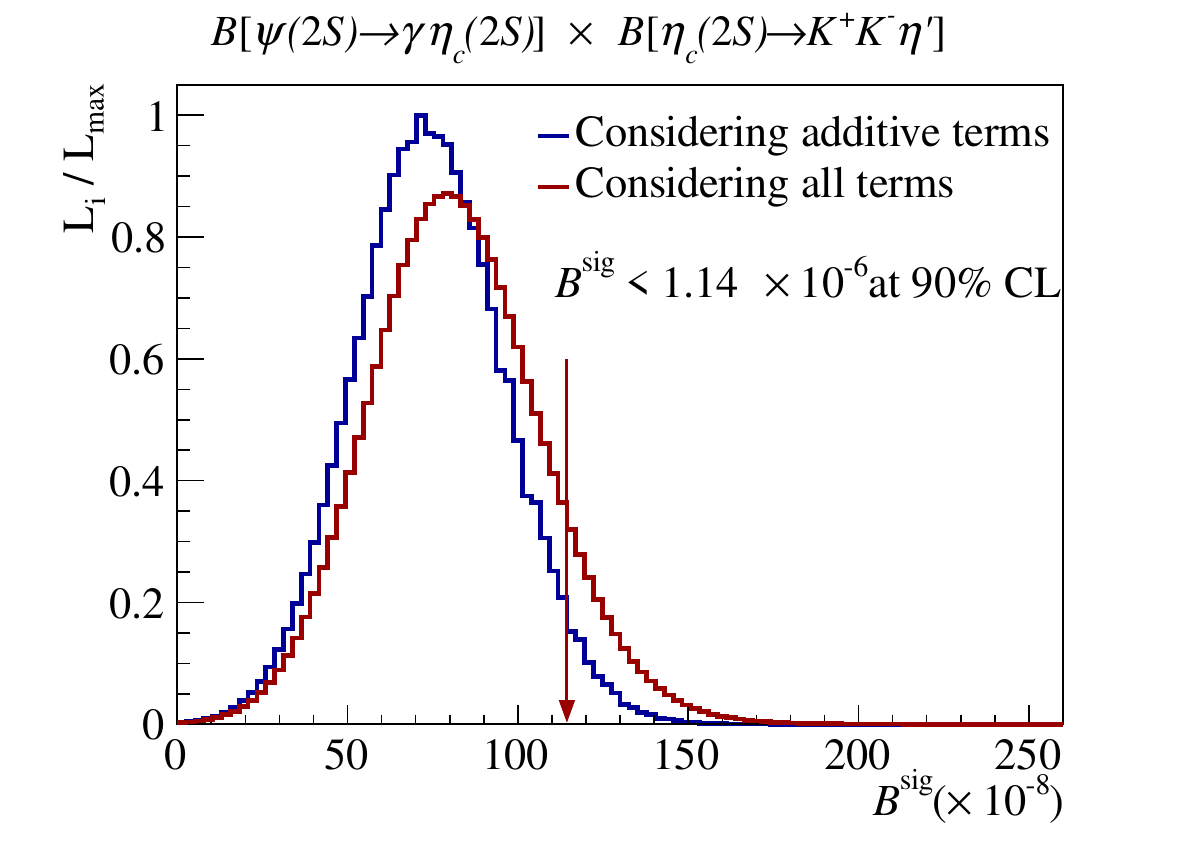}\\
 \caption{The likelihood distribution as a function of the branching fraction of $B(\psip\ra\gamma\etcp \times \etcp\ra\kketap)$. The blue solid line shows the likelihood distribution considering additive uncertainties, and the red solid line shows the likelihood distribution considering all uncertainties. }
 \label{upsys}
\end{figure}

The systematic uncertainty effects on the upper limit of the product branching fraction are considered in two steps. For additive terms, the largest upper limit of the branching fraction from different fit conditions is chosen. The multiplicative systematic uncertainties are incorporated by convolving the likelihood distribution with a Gaussian function~\cite{upper}. The likelihood distributions are shown in Fig.~\ref{upsys}. The upper limit on the product branching fraction of $B[\psip\ra\gamma\etcp] \times B[\etcp\ra\kketap]$ at 90\% C.L is determined to be $1.14\times 10^{-6}$.

\section{SUMMARY}
Using $(2.712\pm0.014)\times10^{9}$ $\psip$ events collected by the BESIII detector, we have searched for the decay $\detacp$ in $\psip$ radiative decays with the two decay modes $\dpipigam$ and $\dpipieta$. 
The branching fraction of $B[\psip\ra\gamma\etcp] \times B[\etcp\ra\kketap]$ is $\rm (6.37\pm2.39(stat.)\pm1.24(syst.))\times 10^{-7}$. 
The branching fraction of $\detacp$ is measured to be $\rm (12.24\pm4.60(stat.)\pm2.37(syst.)\pm4.68(extr.))\times 10^{-4}$, using the branching fraction of $\toetacp=(5.2\pm0.3\pm0.5^{+1.9}_{-1.4})\times10^{-4}$~\cite{psiptogametacp}. The statistical significance of $\detacp$ is 3.1$\sigma$. 
The upper limit on the product branching fraction of $B(\psip\ra\gamma\etcp \times \etcp\ra\kketap)$ at 90\% C.L is determined to be $1.14\times 10^{-6}$. 

The branching fractions of $\chi_{c1,2}\ra\kketap$ are determined to be 
$\rm (8.47\pm0.09(stat.)\pm0.47(syst.))\times 10^{-4}$ and $\rm (1.53\pm0.04(stat.)\pm0.08(syst.))\times 10^{-4}$ for $\chico$ and $\chict$. These results are consistent with the previous measurement~\cite{pdg} (considering the correlation between this and the previous measurement), but with significantly improved precision. 

\begin{figure}[h] 
 \setlength{\abovecaptionskip}{0 cm}  
 \setlength{\belowcaptionskip}{0 cm}
 \centering
 \includegraphics[width=0.49\textwidth]{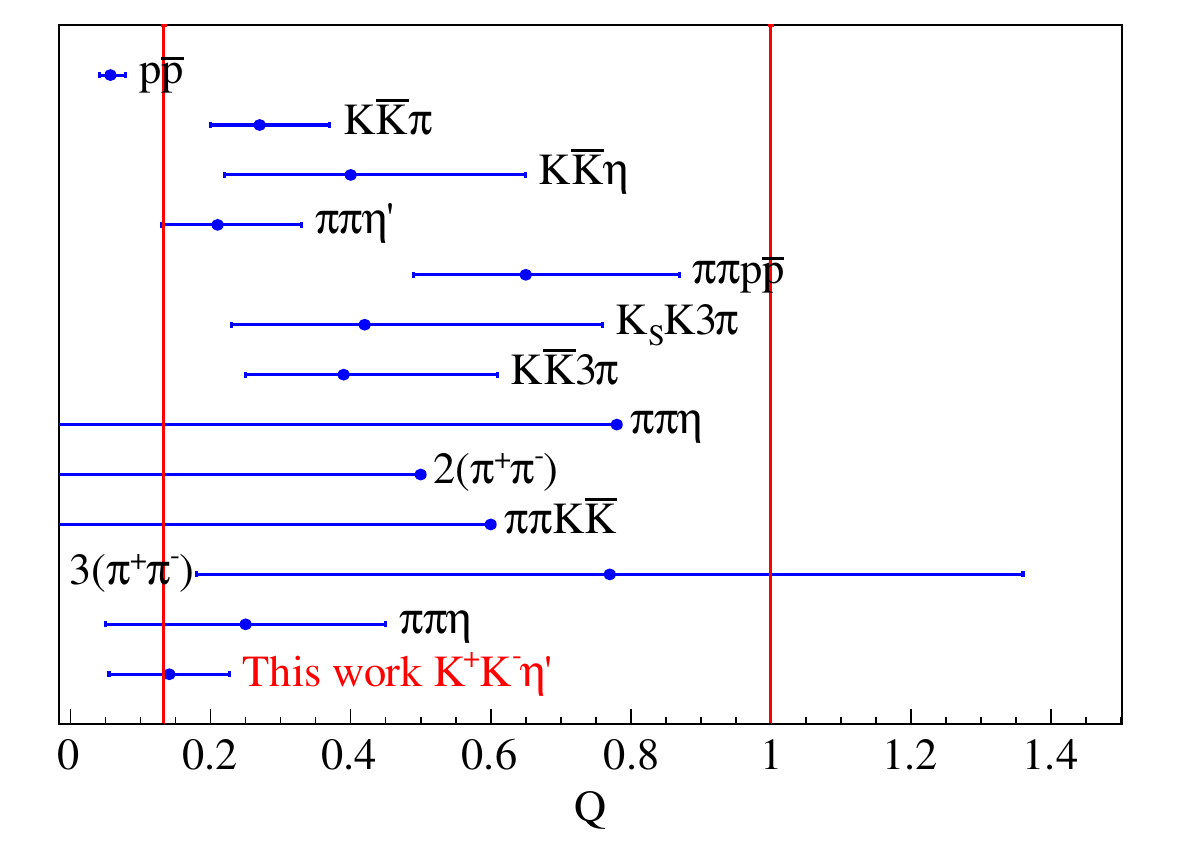}\\
 \caption{The comparison of $\rm Q$ between this study and other hadronic decay modes~\cite{in6:example,lsx,etacp-ppeta}. Dots with error bars are data and the vertical lines show $\rm Q=0.133$ and $\rm Q=1$.  }
 \label{summary_Q}
\end{figure}

Combining our branching fraction of $\detacp$ and that of $\eta_c(1S)\ra\kketap$ from the BaBar Collaboration~\cite{etactokketap}, we determine the branching fraction ratio to be 
\begin{equation}\label{4}
\begin{aligned}  
Q &=\frac{\mathcal{B}(\detacp)}{\mathcal{B}(\eta_{c}(1S)\to\kketap)} 
  =\frac{(12.24\pm6.98)\times10^{-4}}{(0.87\pm0.18)\times10^{-2}},  \\
  &=0.141\pm 0.086,
 \end{aligned}
\end{equation}
where the uncertainty includes the statistical and systematic uncertainties. The correlated uncertainties, e.g. those from the branching fractions of the $\etap$ decays are canceled. 
The comparison of $Q$ from this study and in other hadronic decay modes are shown in Fig.~\ref{summary_Q}. The here obtained $Q$ is closer to the predicted value of 12\% from Ref.~\cite{intro4}, although the uncertainty is large. 
Especially for the decay channels $\etcp\to p\bar{p}$, $\etcp\to K \bar{K}\pi$, $\etcp\to\pi\pi\etap$, and $\etcp\to\pi\pi\eta$, the central values of $Q$ is closer to 12\%. 
More investigations on other decay modes and with improved precision are desired to unveil the underlying mechanism. 

\section{ACKNOWLEDGMENTS}
The BESIII Collaboration thanks the staff of BEPCII and the IHEP computing center for their strong support. This work is supported in part by National Key R\&D Program of China under Contracts Nos. 2020YFA0406300, 2020YFA0406400, 2023YFA1606000; National Natural Science Foundation of China (NSFC) under Contracts Nos. 11635010, 11735014, 11835012, 11935015, 11935016, 11935018, 11961141012, 12025502, 12035009, 12035013, 12061131003, 12192260, 12192261, 12192262, 12192263, 12192264, 12192265, 12221005, 12225509, 12235017, 12150004; Program of Science and Technology Development Plan of Jilin Province of China under Contract No. 20210508047RQ and 20230101021JC; the Chinese Academy of Sciences (CAS) Large-Scale Scientific Facility Program; the CAS Center for Excellence in Particle Physics (CCEPP); Joint Large-Scale Scientific Facility Funds of the NSFC and CAS under Contract No. U2032108, U1832207; CAS Key Research Program of Frontier Sciences under Contracts Nos. QYZDJ-SSW-SLH003, QYZDJ-SSW-SLH040; 100 Talents Program of CAS; The Institute of Nuclear and Particle Physics (INPAC) and Shanghai Key Laboratory for Particle Physics and Cosmology; European Union's Horizon 2020 research and innovation programme under Marie Sklodowska-Curie grant agreement under Contract No. 894790; German Research Foundation DFG under Contracts Nos. 455635585, Collaborative Research Center CRC 1044, FOR5327, GRK 2149; Istituto Nazionale di Fisica Nucleare, Italy; Ministry of Development of Turkey under Contract No. DPT2006K-120470; National Research Foundation of Korea under Contract No. NRF-2022R1A2C1092335; National Science and Technology fund of Mongolia; National Science Research and Innovation Fund (NSRF) via the Program Management Unit for Human Resources \& Institutional Development, Research and Innovation of Thailand under Contract No. B16F640076; Polish National Science Centre under Contract No. 2019/35/O/ST2/02907; The Swedish Research Council; U. S. Department of Energy under Contract No. DE-FG02-05ER41374





\end{document}